\documentclass[11pt,a4paper]{article}
\usepackage[dvips]{graphicx}
\begin{document}
\thispagestyle{empty}
\renewcommand{\baselinestretch}{1.2} 
\small\normalsize
\frenchspacing
\noindent
{\Large \textbf{Schr\"odinger's cat in a realist quantum mechanics}}
\\
\\
{\bf{Arthur Jabs}}
\renewcommand{\baselinestretch}{1}
\small\normalsize
\\
\\
Alumnus, Technical University Berlin. 

\noindent
Vo\ss str. 9, 10117 Berlin, Germany 

\noindent
arthur.jabs@alumni.tu-berlin.de
\\
\\
(29 March 2018)
\newcommand{\rmi}{\mathrm{i}}
\vspace{15pt}

\noindent
{\bf{Abstract.}} There is no paradox with  Schr\"odinger's cat in a realist interpretation because in such an interpretation wave functions which represent the same object at different times are not to be superposed. 

\begin{list}
{\textbf{Keywords:}}
{\setlength{\labelwidth}{2.0cm} 
 \setlength{\leftmargin}{2.2cm}
 \setlength{\labelsep}{0.2cm} }
\item
Schr\"odinger's cat, macroscopic states, superposition
\end{list}

\newcommand{\rmf}{\mathrm{f}}
\newcommand{\rmd}{\mathrm{d}} 
\newcommand{\sbl}{\hspace{1pt}}
\newcommand{\bfitp}{\emph{\boldmath $p$}}
\newcommand{\bfitr}{\emph{\boldmath $r$}}
\newcommand{\bfitsr}{\emph{\footnotesize{\boldmath $r$}}}
\newcommand{\bfitQ}{\emph{\boldmath $Q$}}
\newcommand{\bfitk}{\emph{\boldmath $k$}}
\newcommand{\PSI}[1]{\Psi_{\textrm{\footnotesize{#1}}}}
\newcommand{\PHI}[1]{\Phi_{\textrm{\footnotesize{#1}}}}
\newcommand{\spsi}[1]{\psi_{\textrm{\footnotesize{#1}}}}
\newcommand{\pcop}{P_{\textrm{\footnotesize{Cop}}}}
\newcommand{\pred}{P_{\textrm{\footnotesize{red}}}}
\newcommand{\psis}{\psi_{\rm s}(\bfitr,t)}

\vspace{20pt}
\noindent
In the thought experiment of Schr\"odinger's cat  in the box [1] consider the superposition
\begin{equation}
c_1\psi_1+c_2\psi_2,
\end{equation}
where  $\psi_1$ is the wave function of the alive and $\psi_2$ that of the dead cat. Actually $\psi_1$ may contain the alive cat together with the undecayed radioactive nucleus and $\psi_2$ the dead cat together with the decayed nucleus, but we neglect this because in our present argumentation it is not essential.

The probability of finding, in a measurement (opening the box), an alive cat is $|c_1\psi_1|^2$ and that of the dead cat $|c_2\psi_2|^2$. So there seems to be no objection to write the wave function of the cat inside the closed box as the superposition (1), where $\psi_1$ and $\psi_2$ are probability amplitudes and (1) is the wave function of  neither a dead nor an alive cat but a superposition of both.
\vspace{5pt}

Now take the stand of a realist interpretation, where the particles are wavepackets, $\psi$ represents real matter, not a probability amplitude, and reduction (collapse) is independent of measurement [2], [3]. And should the cat die, this occurs during a definite short time interval around, say, $t_0$, even if nobody takes notice of it. The wavepackets $\psi_1$ and $\psi_2$ here represent  still the same cat (either definitely alive or definitely dead) but \emph{at different times}: $\psi_1$ before $t_0$, $\psi_2$ after $t_0$. The superposition (1)  then superposes the same real object at different times. If we accepted this we would also have to accept the superposition of  an electron wavepacket of today with the same packet of tomorrow. Except for the cat, such a superposition is actually nowhere met in  quantum mechanics, independent of the particular interpretation adopted.

Skeptics should notice that neither is there any superposition of the form (1) in the formulas for transition or scattering amplitudes [4], [5]. And even in the scalar products forming those amplitudes both factors refer to the same time.
\vspace{5pt}

Cases like the Stern-Gerlach experiment are quite different. There, a hydrogen atom passes through an inhomogeneous magnetic field, and its wave function is fanned out into  a superposition of two spatially separated parts, a spin-up part and a spin-down part. The parts reach the screen beyond the magnetic field at the same time, and there the superposition is reduced to one part only. In the Copenhagen view one  then compares the spin up part with the alive and the spin down part with the dead cat, and the reduction in the screen with the opening of the cat's box. In realism the difference to the cat wave functions is, however, that both atom wave functions together represent the atom \emph{at the same time}.
\vspace{5pt}

Thus, although the superposition (1) for the cat is rejected in a realist interpretation, this does not mean that there is no superposition at all of wavepackets representing  mesoscopic or macroscopic objects. The restriction is that these wavepackets must represent something that really exists \emph{at the same time}. This is indeed the case in superpositions of widely separated but localized mesoscopic wavepackets, now sometimes called Schr\"odinger cat states. Examples can be seen in [6]. Ref. [7] reports on molecules of  $1.7\times 10^{-23}$ kg$\, (\approx 10^{4}$ protons) and 5 nm diameter that pass through gratings with slit separation (period) of 266 nm, where interference effects between the parts coming from different slits are indeed observed. Experiments with even larger objects are under way [6], though one difficulty of observing interferences with ever greater objects, such as the cat as a whole, is the increasing importance of environmental decoherence.

\vspace{15pt} 

\noindent
{\textbf{References}} 
\begin{enumerate}
\renewcommand{\labelenumi}{[\arabic{enumi}]}
\hyphenpenalty=1000

\item Schr\"odinger, E.: Die gegenw\"artige Situation in der Quantenmechanik, Die Naturwissenschaften {\bf23}, 807-812, 823-828, 844-849 (1935), the cat appears in \S5, p. 812; English translation in: Proc. Am. Philos. Soc. {\bf124}, 323-338 (1980) and in: Wheeler, J.A. and Zurek, W.H. (eds.): Quantum Theory and Measurement  (Princeton University Press, Princeton, 1983) p. 152-167

\item Jabs, A.: Quantum mechanics in terms of realism, arXiv:quant-ph/9606017 (2017)

\item Jabs, A.: A conjecture concerning determinism, reduction, and measurement in quantum mechanics, arXiv:1204.0614 (2017) (Quantum Stud.: Math. Found. {\bf3} (4), 279-292 (2016))

\item Messiah, A.: Quantum Mechanics (North-Holland Publishing Company, \mbox{Amsterdam}, 1970) p. 725

\item Taylor, J.R.: Scattering Theory (Wiley, New York, 1972) p. 35

\item Arndt, M. and Hornberger, K.: Testing the limits of quantum mechanical superpositions, arXiv:1410.0270 (Nature Physics {\bf10}, 271-277 (2014))

\item Eibenberger, S., Gerlich, S., Arndt, M., Mayor, M., T\"uxen, M.: Matter-wave interference with particles selected from a molecular library with masses exceeding 10 000 amu, arXiv:1310.8343 (Phys. Chem. Chem. Phys. {\bf15}, 14696-14700 (2013))

\end{enumerate}
\bigskip
\hspace{5cm}
------------------------------
\end{document}